\documentstyle[12pt,aaspp4]{article}
\begin{document}

\title{EVIDENCE FOR ABSORPTION DUE TO HIGHLY-IONIZED GAS 
IN THE RADIO-QUIET QUASAR PG~1114+445}

\author {I.M. George \altaffilmark{1,2}, 
K. Nandra\altaffilmark{1,3}, 
A. Laor \altaffilmark{4},
T.J. Turner \altaffilmark{1,2}, 
F. Fiore\altaffilmark{5,6},
H. Netzer \altaffilmark{7},
R.F. Mushotzky\altaffilmark{1}}

\altaffiltext{1}{Laboratory for High Energy Astrophysics, Code 660,
	NASA/Goddard Space Flight Center,
  	Greenbelt, MD 20771}
\altaffiltext{2}{Universities Space Research Association}
\altaffiltext{3}{NAS/NRC Research Associate}
\altaffiltext{4}{Physics Department, Technion, Haifa 32000, Israel}
\altaffiltext{5}{Osservatorio Astronomico di Roma, via 
	dell'Osservatorio 5, Monteporzio-Catone (RM), I-00040, Roma, Italy}
\altaffiltext{6}{SAX Scientific Data Centre, via corcolle 19, 
		I-00131, Roma, Italy}
\altaffiltext{7}{School of Physics and Astronomy and the Wise Observatory,
        The Beverly and Ramond Sackler Faculty of Exact Sciences,
        Tel Aviv University, Tel Aviv 69978, Israel.}

\slugcomment{Accepted for publication in {\em The Astrophysical Journal}
(manusript 36471)}

\begin{abstract}
We present results on the X-ray spectrum of the quasar PG~1114+445 
from an {\it ASCA} observation performed in 1996 June, and a {\it ROSAT} 
observation performed 3 years earlier.
We show good agreement between all the datasets can be obtained if the 
underlying continuum in the 0.2--10~keV band 
is assumed to be a powerlaw (photon index 
$\Gamma \simeq 1.8$) absorbed by photoionized material.
The ionized gas imprints deep absorption edges in the observed spectrum 
$\lesssim 2$~keV due to O{\sc vii} and O{\sc viii}, from which we 
determine its column density ($\sim2\times10^{22}\ {\rm cm^{-2}}$) and 
ionization parameter ($U_X\sim0.1$) to be similar to that observed in 
Seyfert-I galaxies.
Unfortunately these data do not allow any strong constraints to be placed on 
the location, or solid angle subtended by the material at the 
ionizing source.
We also find evidence for absorption in the Fe $K$-shell band in excess 
of that predicted from the lower energy features.
This implies an Fe/O abundance ratio $\sim 10$ times the cosmic value,
or an additional screen of more-highly ionized gas, possibly out-flowing
from the nucleus.
We briefly compare our results with those obtained from other active galaxies.
\end{abstract}


\clearpage
\section{INTRODUCTION}
\label{sec:intro}

A number of {\it ROSAT} and {\it ASCA} X-ray observations of low luminosity 
Active Galactic Nuclei (AGN) have shown clear evidence for absorption by 
highly-ionized gas along the line-of-sight (e.g. Nandra \& Pounds 1992; 
Turner et al 1993; Reynolds 1997; George et al 1997 and references therein). 
The presence of such gas is revealed primarily by the O{\sc vii} and 
O{\sc viii} absorption edges 
imprinted on the underlying continuum.
Recent studies show that the ionized gas generally has a relatively large 
column density (corresponding to effective hydrogen column densities 
$N_H \sim 10^{21}$--$10^{23} {\rm cm}^{-2}$, although there are 
obvious selection effects due the bandpasses and sensitivities of the 
instruments), and exists in $\sim 50$--75\% of Seyfert-I galaxies 
(Reynolds 1997; George et al 1997, thereafter G97).

The situation in the case of higher-luminosity AGN is less clear.
Recent studies have shown that a large fraction of radio-loud quasars
(RLQs) show evidence for intrinsic absorption 
(Cappi et al 1997).
The ionization state of the material generally cannot be determined 
from current data, although there is strong evidence that it is highly 
ionized in at least two cases 
(3C~351, Fiore et al 1993; 3C~273, Grandi et al 1997).
Interestingly, intrinsic absorption appears to be far less common 
in radio-quiet quasars (RQQs), which
are thought to make up $\sim$90\% of total quasar population
(Laor et al 1997; Fiore et al 1997).
Indeed there is only two cases documented to-date:
MR~2251-178 (Halpern 1984, Pan, Stewart \& Pounds 1990, Mineo \& Stewart 1993)
and IRAS~13349+2438 (Brandt, Fabian \& Pounds 1996).
Such a difference between the two classes could be taken to mean that 
gas does not exist along the line--of--sight in RQQs. Alternatively, 
a substantial column density of gas could be present, but so highly ionized 
to be transparent in the X-ray band, or that the absorption features imprinted
by such gas could be swamped by continuum photons 
arriving via another (transparent) travel-path. 

PG~1114+445 ($z=0.144$; Schmidt \& Green 1983) is a RQQ
with $L_{IR} \sim L_{UV} \simeq 10^{45} \ {\rm erg\ s^{-1}}$,
lying in a direction of relatively low Galactic line-of-sight 
column density ($N_{HI}^{Gal} = 1.94\times10^{20}\ {\rm cm^{-2}}$ 
from the 21cm measurements of Murphy et al 1996).  
These characteristics led to the inclusion of the source as one of 
23 quasars forming the complete sample of the Bright Quasar Survey, studied
by the Position Sensitive Proportional Counter (PSPC) on 
{\it ROSAT} (Laor et al 1997).
Interestingly, PG~1114+445 was one of the few objects whose PSPC spectrum 
could not be adequately modelled by a single power-law and 
the only object for which there was 
strong evidence for absorption by ionized material
(Laor et al. 1994).
Further evidence for absorption by ionized material comes from a 
recent {\it HST} observation which has revealed UV absorption lines 
at C{\sc iv}, N{\sc v} and Ly$\alpha$ (Mathur 1997).
Here we present the results from a follow-up {\it ASCA} observation of
PG~1114+445, along with a reanalysis of the PSPC data.

\section{OBSERVATIONS AND DATA REDUCTION}

The new observation reported here was performed using {\it ASCA} on 
1996 May 06--07.
The {\it ASCA} satellite (Makishima et al. 1996) consists of four identical, 
co-aligned X-ray telescopes (XRTs; Serlemitsos et al. 1995). Two solid-state 
imaging spectrometers (SIS0 and SIS1), each consisting of four CCD chips, sit
at the focus of two of the XRTs, and provide coverage over the
$\sim$0.4--10~keV band (Burke et al. 1994). Two gas imaging spectrometers
(known as GIS2 and GIS3) sit at the focus of the other two XRTs,
and provide coverage over the $\sim$0.8--10~keV band (Ohashi et al. 1996
and references therein).
The observation reported here was carried out in \verb+1-CCD+ mode 
with the target in the nominal pointing position. 
The data collected in \verb+FAINT+ and \verb+BRIGHT+ telemetry modes were 
combined.
We employed the same data selection criteria and data analysis methods 
for the {\it ASCA} data 
as presented in Nandra et al (1997a) using the {\tt FTOOLS/XSELECT} package
(v3.5 \& 3.6).
Combined, these gave rise to $5$--$7\times10^4$~s of useful data in each of the 
detectors, from which we derive mean source count rates of 
$(4.5\pm0.1)\times10^{-2}\ {\rm ct\ s^{-1}}$
and 
$(3.4\pm0.1)\times10^{-2}\ {\rm ct\ s^{-1}}$
in the SIS0 and GIS2 detectors respectively, 
and similar rates in SIS1 and GIS3.
The total number of source photons detected in each instrument 
was $2$--$3\times10^3$.
We find no significant variability on 
timescales $\lesssim90$~min, and only marginal evidence
changes in flux on longer timescales.

We have also independently analysed the {\it ROSAT} PSPC 
observation of PG~1114+445 presented by Laor et al. (1994), having 
obtained the data from the archive.
This observation was carried out on 1993 Jun 06--07, and using standard 
data selection and reduction techniques gives rise to 
$7\times10^3$~s of useful data and a mean source count rate of
$(10.2\pm0.4)\times10^{-2}\ {\rm ct\ s^{-1}}$.
No significant variability was observed.

A number of serendipitous sources are apparent within the 
field of view of {\it ROSAT}, at least one of which is 
also detected in the {\it ASCA} GIS, 
but at sufficiently large angular distances not to 
affect the analysis of PG~1114+445 presented here.

\section{SPECTRAL RESULTS}
\label{Sec:results}

Given the lack of strong variability during the observations, mean spectra 
were constructed for each detector, and it is the results from 
these that we concentrate on for the rest of the paper.
As is common practice, in all cases we fit the data from each detector
simultaneously, but allowing the normalization of each to be a free parameter 
to allow for differences in the sizes of the extraction cell used, 
along with any discrepancies in the absolute flux 
calibration of the individual telescope/detector systems.
The spectral analysis was performed using \verb+XSPEC+ (v9.01; Arnaud 1996). 
Appropriate detector redistribution matrices were used (those released 
1994 Nov 09 and 1995 Mar 06 for the {\it ASCA} SIS and GIS respectively,
and \verb+pspcb_gain2_256.rmf+ for the {\it ROSAT} PSPC).
The effective 
area appropriate for each dataset was calculated using 
\verb+ascaarf+\footnote{so as to include the Gaussian 'fudge', but 
not the filter 'fudge' in the case of the SIS datasets. We have also 
compared our results to those obtained using several developmental versions 
of  \verb+ascaarf+ (up to and including v2.64), but find no significant 
differences.}(v2.5) for the {\it ASCA}
data, and \verb+pcarf+(v2.1.0) for the PSPC data.
In the case of the SIS data, we 
have restricted our spectral analysis to energies $\geq 0.6$~keV
due to residual uncertainties in the calibration of the SIS/XRT instrument
(see Dotani et al. 1996 and the information provided by the 
{\it ASCA} GOF at NASA/GSFC\footnote{
see \verb+http://heasarc.gsfc.nasa.gov/docs/asca/cal\_probs.html+
for up-to-date information}).
However, whilst the calibration is suspect at these energies, it is considered
unlikely to be in error by $\gtrsim20$\%. Thus, following G97 we have also
calculated the weighted mean of the data/model residuals
($\overline{R_{0.6}}$), when the best-fitting model is extrapolated below
0.6~keV. The rationale behind this parameter is that it allows us to identify
models, which are deemed acceptable $>0.6$~keV, but in which the extrapolation
to energies $<0.6$~keV is inconsistent with the suspected size of the
calibrations uncertainties.

The raw spectra extracted for each detector were grouped such that each 
resultant channel had at least 20 counts per bin, permitting us to use 
$\chi^2$ minimization during the spectral analysis. 
This results in a total of 511 spectral bins for the combined {\it ASCA}
detectors, and 37 spectral bins for the PSPC data.
In passing, we note that in the 0.6-0.8 keV band these grouped spectral bins 
are each typically 0.06~keV wide (for both the SIS and PSPC datasets).
This is comparable to the spectral resolution of the SIS at the epoch of 
the observations (FWHM $\simeq 0.09$~keV) and far smaller than the 
spectral resolution of the PSPC (FWHM $\simeq 0.35$~keV) within this 
energy band.
Appropriate background spectra were extracted from source-free regions of
each detector.

We have compared a number of hypothetical models with the mean 
spectra.
In all models considered here, we have assumed an underlying
continuum represented by a single power-law (of photon index $\Gamma$) 
throughout the 0.1--10~keV band (observers frame).
All the models also 
include the effects of absorption by neutral material at $z=0$ 
parameterized by an effective hydrogen column density $N_{H,0}$ 
(assuming the abundances and cross-sections of 
Morrison \& McCammon 1983), which can be allowed to vary during
the spectral analysis (but constrained to be $\geq N_{HI}^{Gal}$).
Most models include an additional column density $N_{H,z}$ 
of absorbing material at the redshift of the source.
Errors are quoted at 68\% confidence for the appropriate number of 
interesting parameters (where all free parameters are considered interesting
except for the absolute normalization of the model).

\subsection{Analysis of the overall continuum}
\label{Sec:wabs}

\subsubsection{Fits to the {\it ASCA} data}

We find a simple power-law fit to the {\it ASCA} data in the 0.6--10~keV band 
confirms the presence of substantial absorption. If the absorbing material is 
assumed to be neutral we find 
$N_{H,0} \simeq 6\times10^{21}\ {\rm cm^{-2}}$
or 
$N_{H,z} \simeq 7\times10^{21}\ {\rm cm^{-2}}$
depending whether the material is assumed to be local or at the redshift of 
the quasar. Both fits gave a photon index $\Gamma \simeq 1.6$ and statistically 
acceptable fits (with reduced $\chi^2$ value of $\chi^2_{\nu} = 1.0$).
However the extrapolation of both best-fitting models predicts fewer counts 
$<0.6$~keV than observed by a factor $\overline{R_{0.6}}\sim14$, and give 
rise to an increase in $\chi^2$-statistic of $\Delta \chi^2_{0.6} \sim 70$
(for 6 additional spectral bins). Such a discrepancy is far greater than can 
be accounted for remaining uncertainties in the instrumental calibration.
As we shall see below, these solutions are artifacts of the presence of
absorption by ionized gas.

Numerous elements have photoelectric absorption edges with threshold energies 
within the 0.6--1.0 keV band. However, for cosmic abundances, the opacity 
in this band is dominated by $K$-shell absorption by C, O and Ne, and 
$L$-shell absorption by Fe
(e.g. Morrison \& McCammon 1983). Given the redshift of PG~1114+445, the most 
significant edges likely to be observable in {\it ASCA} spectra are those due 
to O{\sc vii} and O{\sc viii}, with rest--frame energies of 739 and 871~eV 
respectively. Including such edges in our analysis (fixing 
$N_{H,0} = N_{H,0}^{Gal}$ and $N_{H,z} = 0$), we find optical depths in 
these species of $\tau$(O{\sc vii})$\simeq 2.5$ and 
$\tau$(O{\sc viii})$\simeq 1.1$. Such a model can be considered a crude 
parameterization of the absorption features imprinted by ionized gas along 
the line--of--sight to the nucleus. However, at such large optical depths
absorption by other abundant elements is likely to be important also, 
hence a more detailed treatment is required.
Here 
we consider the case where the material is ionized by the central 
continuum, and employ theoretical spectra generated using 
the photoionization code {\tt ION} (Netzer 1993, 1996, version {\tt ION95}). 
The ionization state of the gas is parameterized by the 
'X-ray ionization parameter' $U_X$
(Netzer 1996; where the ionizing radiation field is determined 
over the 0.1--10~keV band).
Further details on the assumptions made in the {\tt ION} 
models and the method by which they were included in the 
spectral analysis can be found in G97.

We find that including the absorption by photoionized gas 
provides an excellent description 
of the {\it ASCA} data (with $N_{H,0}=N_{HI}^{Gal}$),
giving
$\Gamma \simeq 1.8$, 
$N_{H,z} \simeq 2\times10^{22}\ {\rm cm^{-2}}$
and 
$U_X \simeq 0.1$ (Table~1, line 1).
The addition of $U_X$ as an additional free parameter 
is significant at $>$99.9\% confidence ($F$-statistic of 33.7).
Furthermore this model 
gives far better agreement with the {\it ASCA} data 
$<0.6$~keV ($\overline{R_{0.6}}\sim0.8$, consistent with 
the current uncertainties in the instrumental calibration).
When such a model is assumed, we find
$N_{H,0}$ consistent with $N_{HI}^{Gal}$, with an
upper limit to any such excess absorption 
of $\Delta N_{H,0} \simeq 3\times10^{21}\ {\rm cm^{-2}}$
(Table~1, line 2).
The best-fitting model spectrum, along with the observed data/model 
ratios (for the $N_{H,0}=N_{HI}^{Gal}$ case) are shown in Fig~1, 
and the $\chi^2$-contours in $N_{H,z}$--$U_X$ space
shown (solid lines) in Fig~2.
As can be seen from Fig.~1, for the best fitting values of 
$N_{H,z}$ and $U_X$ the intervening gas imprints a 
series of absorption edges on the underlying continuum, dominated 
by C{\sc vi}, O{\sc vii} and O{\sc viii}.
However, it should be noted that the ionization level of the gas is such that 
it becomes 
increasingly transparent as one moves to lower energies below the 
O{\sc vii} edge (0.65~keV in the observer's frame).
It is this behaviour that leads to a better agreement between the extrapolated 
model and the observed SIS data $<0.6$~keV, and we suggest, with the 
PSPC data (see below).
We have considered models in which the ionized absorber 
attenuates only a fraction of the underlying continuum. However in all 
cases we find a covering-fraction consistent with 100\%, 
thus such models are not considered 
further. 
Stringent constraints cannot be placed on the redshift of the 
ionized gas, $z_{abs}$, using the current 
data. However we do find $z_{abs}> 0.103$ at 90\% confidence, 
consistent with the systemic redshift of the host galaxy, and 
limiting the ionized gas to be in an 
outflow/wind of velocity $ \lesssim 10^{4}\ {\rm km\ s^{-1}}$
relative to the nucleus\footnote{It should be noted that 
the spectral energy resolution of the {\it ASCA} SIS 
($\Delta E/E \simeq 0.14$ at the epoch of the observations) 
prevents the 
velocity distribution of the ionized gas being constrained within 
$\sigma_v \lesssim 2\times 10^{4}\ {\rm km\ s^{-1}}$}.

With the assumption
that our line-of-sight is not uniquely privileged, we have also tested 
whether
any useful constraints can be obtained on the solid angle, $\Omega$, subtended
by such gas at the central continuum source from its expected
emission/scattering spectrum. However, the inclusion of such a 
component in the spectral analysis (where $N_{H,z}$ and $U_X$ 
of the emitting gas is the same as those for the 
absorbing gas) is inconclusive. We find an upper limit on the strength of 
such a component is that expected from a full shell of such material
($\Omega = 4\pi$) assuming it was irradiated by the continuum source 
with a luminosity a factor $\lesssim 8$ times greater than that derived
from the observations.
This is hardly surprising given that by far the bulk of the emission expected 
from ionized gas in this region of $N_{H,z}$, $U_X$ parameter--space
occurs below the O{\sc vii} edge, and hence $<0.6$~keV in the observer's frame.

\subsubsection{Fits to the {\it ROSAT} data}

Considering the PSPC data alone, we confirm the results of Laor et al (1994)
that a single powerlaw does not provide an adequate description of the 
data (giving $\chi^2/dof = 59.5/35$), 
but that a statistically satisfactory fit can be obtained with the 
addition of a deep absorption edge. From our analysis, fixing 
$N_{H,0}$ at $N_{HI}^{Gal}$, we find
a best fitting solution with an edge of optical depth 
$\tau=3.5^{+1.7}_{-1.3}$ 
at an energy $E_z = 0.77^{+0.05}_{-0.06}$~keV 
(in the rest--frame of the quasar) imprinted on an 
underlying continuum with $\Gamma = 1.96^{+0.17}_{-0.17}$
and $\chi^2/dof = 23.1/33$.
These values are
in agreement with those found by Laor et al (1994; 
$\tau \simeq 3$ at $E_z \simeq 0.76$~keV), and with a blend of 
the O{\sc vii} and O{\sc viii} edges found in the analysis of the 
{\it ASCA} data.
Repeating the analysis, but assuming the photoionization models described 
above (with $N_{H,0} = N_{HI}^{Gal}$), we find
$N_{H,z}   = 37^{+41}_{-22}\times10^{21}\ {\rm cm^{-2}}$, 
$U_X     = 0.24^{+0.11}_{-0.14}$, 
and $\Gamma   = 1.7^{+0.3}_{-0.3}$
with $\chi^2/dof = 22.7/33$.
These values are 
in good agreement with those obtained from the {\it ASCA} data above.
However, despite the fact that the PSPC provides some data
at energies $<0.6$~keV, 
the low signal-to-nose ratio of these data along with the 
relatively poor spectral resolution of the detector lead to 
few additional constraints on the state of the ionized gas 
being provided by the PSPC data alone.

\subsubsection{Joint analysis of the {\it ASCA} and {\it ROSAT} data}

Here we present the results 
from a joint analysis of both the {\it ASCA} and {\it ROSAT} datasets
which may give better constraints on the properties of the 
ionized gas.
As previously, 
the same spectral model is assumed for the data from each mission, but the 
normalization (only) of the underlying powerlaw is allowed to vary 
in order to allow for any intensity 
variations in the source over the 3 years between the observations.

As expected, statistically acceptable results are obtained with the 
photoionization model
(Table~1 lines 3 \& 4), with 
the flux during the {\it ROSAT} observation being a factor 
$\sim90$\% of that at the time 
of the {\it ASCA} observations. This is close to the accuracy of the 
the absolute flux calibration between the two instruments, so is 
not significant.
It can be seen that the best-fitting parameters are also consistent
with those derived when the {\it ASCA} and {\it ROSAT} data are 
considered individually.
The $\chi^2$-contours in $N_{H,z}$--$U_X$ space assuming 
$N_{H,0} = N_{H,0}^{Gal}$ are also shown (dashed curves) in Fig~2.
Allowing the redshift of the ionized gas to vary during the analysis
we find $ 0.115 \leq z_{abs} \leq 0.217 $ (at 90\% confidence), 
again consistent with the systemic redshift of the host galaxy, and
limiting the ionized gas to have a velocity 
$\lesssim \pm 2\times 10^{4}\ {\rm km\ s^{-1}}$ with respect to the 
nucleus.
In passing, we note however, that the best-fitting models for this
combined analysis do consistently overpredict the number of counts 
observed in SIS $<0.6$~keV ($\overline{R_{0.6}} \sim 0.5$--$0.7$).
This may simply be due to the residual uncertainties in the calibration 
of the SIS/XRT instrument at these energies, or be an artifact of 
fitting non-simultaneous datasets with different signal-to-noise ratios
from instruments with very different energy resolutions.
The strength of any emission component is restricted to be that 
from a full shell of such material assuming it was irradiated by the 
continuum source 
with a luminosity a factor $\lesssim 1.5$ times that derived from the 
observations.

\subsection{Results from the Fe $K$-shell band}
\label{Sec:fe-band}

The analysis presented in \S\ref{Sec:wabs} did not take into account 
the potential presence of features in the 6--9~keV band 
(in the rest--frame of the quasar) due to Fe $K$-shell transitions.
Indeed, the data/model residuals shown in Fig.~1 do show a
deficit in this energy range.
This absorption feature appears to 
be present in all 4 {\it ASCA} detectors, and can be modelled with 
an absorption edge of energy $E_z^{abs}$ (in the rest--frame of the quasar) and 
optical depth 
$\tau$ (where $\tau \propto (E/E_z^{abs})^{-3}$ for energies $E \geq E_z^{abs}$
and zero elsewhere). 
Considering the {\it ASCA} data alone and fixing 
$N_{H,0} = N_{H,0}^{Gal}$, 
the addition of these two parameters results in an improvement 
of $\Delta \chi^2 = 9.2$, significant at $> 95$\% confidence 
($F$-statistic = 4.9).
We find $\tau = 0.35^{+0.38}_{-0.29}$ 
and $E_z^{abs} = 7.25^{+0.42}_{-0.48}$~keV.
The confidence contours in the $E_z^{abs}$--$\tau$ plane 
are shown in Fig.~3a, along with the corresponding 
edge energies for Fe $K$-shell absorption. 
It can be seen that the edge is consistent with
Fe{\sc i}--Fe{\sc xix} at 90\% confidence for $\tau > 0.1$, 
assuming absorbing material in the rest--frame of the quasar.
The best-fitting $\tau$ corresponds to an equivalent hydrogen 
column density of $\sim 10^{19}/A_{\rm Fe} \ {\rm cm^{-2}}$ 
where $A_{\rm Fe}$ is the abundance of Fe relative to hydrogen.
Thus, if the Fe feature is produced in the same material 
as is responsible for the O features observed $\lesssim 1$~keV 
discussed above ($N_{H,z} \simeq 2\times10^{22}\ {\rm cm^{-2}}$), 
then we require $A_{\rm Fe} \sim 5\times10^{-4}$, a factor 
$\sim 10$ greater than the most recent estimations of the 
'cosmic' abundance (Anders \& Grevesse 1989).

We note that a yet superior fit is obtained if the absorption feature is 
assumed to be 'notch'-shaped (a saturated line with vertical sides, and 
$\tau = \infty$ within), giving $\Delta \chi^2 = 18$ compared to the model 
with no absorption. The best-fitting notch has an equivalent width of 
$343^{+214}_{-205}$~eV (equivalent to one rebinned spectral bin)
and is centered at an energy $7.76^{+0.16}_{-0.24}$~keV
(in the rest--frame of the quasar; Fig.3b)
Assuming this is indeed material in the rest--frame of the quasar, 
such an energy is consistent 
with resonance scattering by Ni $K\alpha$ ($<$Ni{\sc xxviii})
and Fe $K\beta$ ($>$Fe{\sc xx}).
Although both families of interactions include transitions with relatively 
large oscillator strengths ($\sim 0.7$--0.8), both interpretations are 
problematic.
In the former case, the cosmic abundance of Ni is far smaller than that for Fe
($A_{\rm Ni}/A_{\rm Fe} \simeq 0.04$) making such an interpretation 
unlikely due to the lack of corresponding features due to Fe at lower energies.
In the latter case, resonance scattering by Fe $K\alpha$ will dominate that 
due to Fe $K\beta$ for ionization states $>$Fe{\sc xvi}, again making such an 
interpretation unlikely due to the lack such a feature in 6.4--6.9~keV band.
An alternative explanation of the notch is that it is due to 
resonance scattering by Fe $K\alpha$, but that the material is
out--flowing ($\sim 0.1$c for Fe{\sc xxv})
with respect to the rest-frame of the quasar.
However, the instrumental and cosmic backgrounds start to become a noticeable 
fraction of the observed signal from PG~1114+445 in this observation above 
$\sim 6$~keV. Given that both the SIS and GIS detectors contain background 
features in this region (e.g. Gendreau 1994; Makishima et al. 1996), we 
consider the form of the absorption to require confirmation before more 
detailed interpretations can be made.

Emission features due to Fe $K$-shell processes 
are common in low luminosity AGN (e.g. Nandra \& Pounds 1994), 
and
have been claimed in high luminosity AGN (e.g. Williams et al. 1992; 
Nandra et al 1996). With such a deep Fe absorption in
our spectrum, we might expect to observe the associated
K$\alpha$ emission. For the ionization levels suggested by the above fits 
including an edge ($<$Fe{\sc xx})
the emission line energy should be close to 6.4~keV. We have tested for the
presence of such a line by adding a narrow Gaussian line to the model
at that energy. Although a significant improvement
is not obtained, the equivalent width $W_{K\alpha} = 60^{+120}_{-60}$~eV 
is consistent with the predictions of photoionization models assuming 
full coverage of the column density implied
by the edge fits. If the emission line is broad, as observed
in Seyfert 1 galaxies, then the upper limit to 
$W_{K\alpha}$ is larger by a factor $\sim 5$.
 
\section{DISCUSSION AND CONCLUSIONS}

In the previous section we have shown that both the {\it ASCA} and 
{\it ROSAT} spectra from PG~1114+445 are consistent with
a single powerlaw continuum with $\Gamma \simeq 1.8$ absorbed 
by a column density of $\simeq 2\times10^{22}\ {\rm cm^{-2}}$
of photoionized gas $U_X \simeq 0.1$.
We suggests this offers the most plausible explanation of the
the X-ray spectrum of this source, making it only the 
fifth quasar (alongside MR~2251-178, 3C~351, 3C~273 and
IRAS~13349+2438)
in which ionized material has been detected along the line--of--sight.
After correcting for absorption (i.e. setting $N_{H,0} = N_{H,z} = 0$),
the X-ray luminosities\footnote{assuming 
$H_0 =  50\ {\rm km\ s^{-1}\ Mpc^{-1}}$ and $q_0 = 0.5$}
are $L_X \simeq 5.7\times 10^{44}\ {\rm erg\ s^{-1}}$ 
and $2.7\times 10^{44}\ {\rm erg\ s^{-1}}$
over the 0.1--10 keV and 2--10~keV bands
(source-frame) respectively.
This is comparable to the higher-luminosity examples of the 
sources considered by G97.
Unfortunately these data do not allow any strong constraints to be placed on 
the location, or solid angle subtended by the ionized material at the 
ionizing source.

Clearly a comparison between the properties of the material dominating 
the absorption in the X-ray band of PG~1114+445 
with that responsible for the absorption features seen in the 
UV might be revealing.
To this end, in Table~2 we list the column densities of the 
abundant Li-like ions predicted from our best-fit model from 
\S3.1.
It should be stressed that such predictions 
are very sensitive to form of the ionizing continuum in the XUV band,
and of course the elemental abundances.
The values quoted in Table~2 assume the 'weak IR' continuum described 
by Netzer (1996), namely a {\it photon} index $\Gamma_{o} = 1.5$ in the 
optical/UV band from 1.6--40.8~eV, an index $\Gamma = 1.9$ in the X-ray band
from 0.2--50~keV, and a powerlaw continuum in the XUV band 
connecting the fluxes at 40.8~eV and
0.2~keV, such that the ratio of the fluxes at 250~nm and 2~keV
is $f_{250}/f_{\rm 2keV} = 8.1\times10^{3}$
(corresponding to an optical--to--X-ray {\it energy} index
$\alpha_{ox} = 1.5$).
Following
Netzer (1996), we assume undepleted 'cosmic abundances' (with
(He, C, N, O, Ne, Mg, Si, S, Fe)/H =
($10^3$, 3.7, 1.1, 8, 1.1, 0.37, 0.35, 0.16, 0.4)/$10^{-4}$).
For comparison, the column densities in the bound--free transitions
in O{\sc vii} \& O{\sc viii} are 
$9.6\times10^{18}\ {\rm cm^{-2}}$ \& $5.1\times10^{18}\ {\rm cm^{-2}}$
respectively.
In passing we note that given the cross-sections for these transitions
($2.4\times10^{-19}\ {\rm cm^2}$ \& 
$9.9\times10^{-20}\ {\rm cm^2}$ respectively), the predicted 
optical-depth of O{\sc vii} ($\tau \simeq 2.5$)
is in good agreement with that 
obtained in \S3.1.1 when the data is modelled as a powerlaw plus
O{\sc vii} \& O{\sc viii} absorption edges.
However, the observed optical-depth of 
O{\sc viii} ($\tau\simeq 1.1$) is a factor $\sim$2 
larger than that predicted in our full photoionization calculations,
primarily as a result of the modelled optical-depth of
O{\sc viii} having to include the additional 
opacity due to Fe $L$-shell and Ne{\sc ix} $K$-shell transitions 
during the fitting process.

The parameters for the ionized gas obtained here for PG~1114+445 
are very similar to those found for Seyfert-I galaxies. 
G97 
find the ionization parameter typically clusters around 
$U_X \sim 0.1$ with column densities in the 
range $N_{H,z} \sim 10^{21}$--$10^{23}\ {\rm cm^{-2}}$.
The underlying spectral index also lies in the range found in 
Seyfert-I galaxies.
The underlying continuum could not be determined unambiguously from the 
{\it ROSAT} PSPC data from the RLQ 3C~351
($z = 0.371$, $L_X \simeq 3\times 10^{45}\ {\rm erg\ s^{-1}}$).
However the fits including absorption by ionized material reported by 
Fiore et al (1993) indicate a similar column density 
($N_{H,z} \sim 1$--$4\times10^{22}\ {\rm cm^{-2}}$)
to that found in PG~1114+445 and Seyfert-Is, but that the gas is 
less ionized ($U_X \sim $few$\times10^{-2}$, after conversion of the quoted 
ionization parameter to that over the 0.1--10~keV band used here).
Nandra et al (1997b) have recently reported the results from {\it ASCA} 
observations of the RQQ MR~2251-178 
($z = 0.068$, $L_X \simeq 2\times 10^{45}\ {\rm erg\ s^{-1}}$),
obtaining a lower column density 
($N_{H,z} \sim 2\times10^{21}\ {\rm cm^{-2}}$)
and $U_X \sim 0.07$.
Brandt, Fabian \& Pounds (1996) have found evidence for absorption 
by ionized oygen in {\it ROSAT} PSPC data from 
IRAS~13349+2438
($z=0.107$, $L_X \sim 10^{45}$--$10^{46}\ {\rm erg\ s^{-1}}$).
Unfortunately the properties of the ionized material could not be 
well constrained by those data 
($N_{H,z} \sim$few$\times10^{21}$--$10^{24}\ {\rm cm^{-2}}$,
$U_X \sim 0.2$--$0.5$).
However, the lack of significant 
absorption by {\it neutral} material in the PSPC data along with 
strong evdience for dust at other wavebands led Brandt et al to suggest 
that the dust may be embedded within the ionized material.

A particularly interesting result of our analysis is the detection
of an absorption feature within the Fe $K$-shell band (\S 3.2). 
The strength of this feature 
implies either an Fe/O abundance ratio $\sim 10$ times cosmic, or a second
zone of highly ionized material, perhaps outflowing from the nucleus.
The feature is of similar energy and depth as those observed in 
Seyfert-I galaxies by {\it Ginga} (Nandra \& Pounds 1994), 
and that possibly detected in an {\it ASCA} observation of the 
RLQ S5~0014+81 (Cappi et al 1997).
Future observations, with higher sensitivity and resolution 
in the 7--9~keV band, are 
required to confirm these features and better determine their nature.

\acknowledgements
We would like to thank Keith Gendreau, Tahir Yaqoob, Tim Kallman,
Smita Mathur, Bev Wills and Fred Hamann for useful discussions.
We acknowledge the financial support of
the Universities Space Research Association (IMG, TJT), 
National Research Council (KN) and Israel Science Foundation (NH).
This research has made use of 
the Simbad database, operated at CDS, Strasbourg, France; of
the NASA/IPAC Extragalactic database, which is operated by
the Jet Propulsion Laboratory, Caltech, under contract with NASA; and
of data obtained through the HEASARC on-line service, provided by
NASA/GSFC.

%%%%%%%%%%%%%%%%%%%%%%%%%%%%%%%%%%%%%%%%%%%%%%%%%%%%%%%%%%%%%%%%%%%%%%
\begin{deluxetable}{lllllll}

\tablecaption{Best-fitting spectral parameters to PG 1114+445
\label{tab:fits}}

\scriptsize
\tablecolumns{7}
\tablewidth{0pt}
\tablehead{
	\colhead{$N_{H,0}$} &
	\colhead{$\Gamma$} &
	\colhead{$N_{H,z}$} &
	\colhead{$\log U_X$}	&
	\colhead{$\chi^2/dof$}	&
        \colhead{$\Delta \chi^2_{0.6}$} &
        \colhead{$\overline{R_{0.6}}$} 
\nl
	\colhead{$(10^{21}\ {\rm cm^{-2}})$}	&
	&
	\colhead{$(10^{21}\ {\rm cm^{-2}})$}	&
	&
	&
	&
}

\startdata
\multicolumn{7}{l}{\it ASCA analysis alone} \nl

	$0.19$ (f)	&
	$1.75^{+0.10}_{-0.10}$  & 
	$19.60^{+4.39}_{-3.82}$	&
	$-1.027^{+0.130}_{-0.162}$	&
	478/504 &	
	10 	&
	\phn0.8	\nl

	$1.25^{+2.01}_{-1.06\ (p)}$	&
	$1.78^{+0.15}_{-0.12}$  & 
	$21.03^{+11.33}_{-6.19}$	&
	$-0.877^{+0.274}_{-0.280}$	&
	475/503 &	
	\phn6 	&
	\phn1.2	\nl

\multicolumn{7}{l}{\it Joint ASCA \& ROSAT analysis} \nl

	$0.22^{+0.07}_{-0.03\ (p)}$	&
	$1.79^{+0.12}_{-0.10}$  & 
	$22.94^{+5.57}_{-4.28}$	&
	$-0.922^{+0.111}_{-0.084}$	&
	508/539 &	
	31	&
	\phn0.6	\nl

	$0.19$ (f)	&
	$1.76^{+0.06}_{-0.07}$  & 
	$21.42^{+3.80}_{-2.64}$	&
	$-0.953^{+0.107}_{-0.105}$	&
	508/540 &	
	23 	&
	\phn0.7	\nl

\tablecomments{\footnotesize
Fits undertaken in the 0.6--10.0~keV and 
0.1--2.0~keV bands in the observer's frame for 
{\it ASCA} and {\it ROSAT} respectively.
$N_{H,0}$ is the neutral column at $z=0$ (constrained to be
$\geq N_{HI}^{Gal} = 1.94\times10^{20}\ {\rm cm^{-2}}$);
$\Gamma$ is the photon index of the underlying powerlaw continuum;
$N_{H,z}$ is the column density of the ionized gas 
at the redshift of the source;
$U_X$ is ionization parameter (see text);
$\Delta \chi^2_{0.6}$ is the increase in the $\chi^2$-statistic, 
$\overline{R_{0.6}}$ the mean data/model ratio
when the best-fitting 
model is extrapolated to the 6 SIS spectral bins $<0.6$~keV.
The errors are at 68\% confidence for 3 (lines 1 \& 3) and
4 (lines 2 \& 4) interesting parameters.
$(f)$ indicates  the parameter was fixed at the specified value, and
$(p)$ that $N_{H,0}$ 'pegged' at $N_{HI}^{Gal}$.
}

\enddata
\end{deluxetable}
\clearpage

%%%%%%%%%%%%%%%%%%%%%%%%%%%%%%%%%%%%%%%%%%%%%%%%%%%%%%%%%%%%%%%%%%%%%%
\begin{deluxetable}{lllllll}

\tablecaption{Derived column densities of Li-like ions for 
$U_{X}=0.1$, $N_{H,z}= 2\times10^{22}\ {\rm cm^{-2}}$
\label{tab:coldens}}

\scriptsize
\tablecolumns{2}
\tablewidth{0pt}
\tablehead{
	\colhead{ion} &
	\colhead{$N_{ion}$}	
\nl
	&
	\colhead{$({\rm cm^{-2}})$}	
}

\startdata
C{\sc iv}	&	$3.1\times10^{14}$	\nl
N{\sc v}	&	$3.9\times10^{15}$      \nl
O{\sc vi}	&	$4.5\times10^{17}$      \nl
Ne{\sc viii}	&	$5.0\times10^{17}$      \nl
Mg{\sc x}	&	$2.9\times10^{17}$      \nl
Si{\sc xii}	&	$1.4\times10^{17}$      \nl
S{\sc xiv}	&	$9.9\times10^{15}$      \nl
Ar{\sc xvi}	&	$1.4\times10^{14}$	

\tablecomments{\footnotesize
It should be stressed that the predicted column densities 
are very sensitive to form of the ionizing continuum in the XUV band,
and of course the elemental abundances (see \S4).
}

\enddata
\end{deluxetable}
\clearpage

\typeout{FIGURES}

\begin{figure}
\plotfiddle{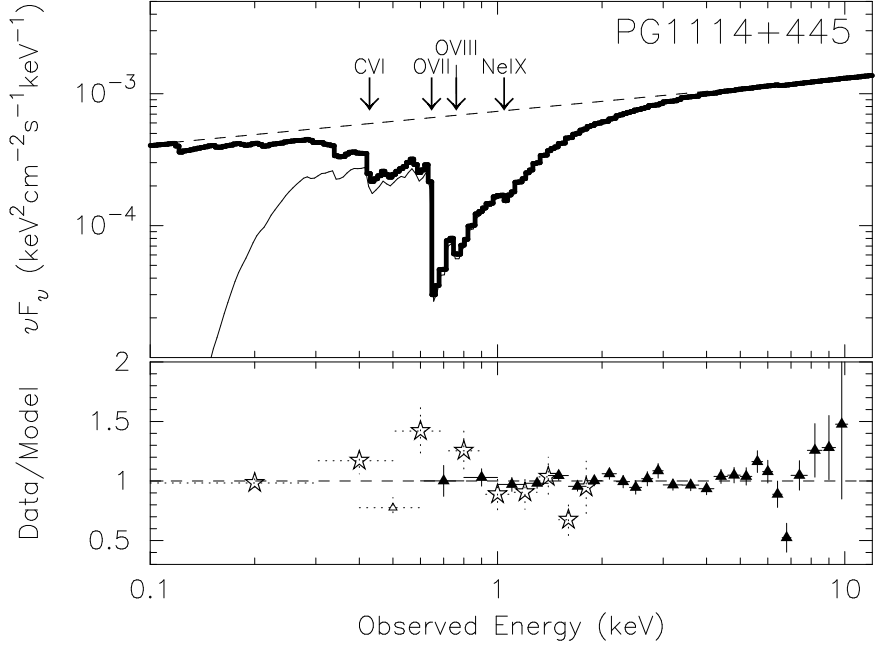}{8cm}{270}{50}{50}{-210}{310}
%\plotone{f1.ps}
\caption{
Upper panel: The preferred ionized-absorber model for PG~1114+445 based on 
the {\it ASCA} observations performed in 1996 Jun. The dashed line 
shows the underlying powerlaw continuum, the bold curve shows the 
best-fitting spectrum (corrected for Galactic absorption
with $N_{HI}^{Gal} = 1.94\times10^{20}\ {\rm cm^{-2}}$)
the faint solid line shows the observed spectrum
(uncorrected for $N_{HI}^{Gal}$).
The deep absorption features due to C{\sc vi}, O{\sc vii} and O{\sc viii}
are clearly visible in the spectrum, as is the fact that the 
ionized gas becomes transparent below $\sim 0.4$~keV.
The preferred model has an ionization parameter
$U_X \simeq 0.1$, and 
an effective hydrogen column denisty 
$N_{H,z} \simeq 2\times10^{22}\ {\rm cm^{-2}}$
assuming abundances relative to hydrogen
of $A_{\rm C}=3.7\times10^{-4}$, 
$A_{\rm O}=8\times10^{-4}$ and
$A_{\rm Ne}=1.1\times10^{-4}$ for C, O and Ne respectively.
Lower panel: the mean data/model ratio from this fit 
(where, for clarity, we show the rebinned, averaged ratios for 
{\it ASCA}).
The filled triangles correspond to the {\it ASCA} data 
used during the 
spectral analysis, whilst the open triangle shows the ratio when 
the best-fitting model is extrapolated $<0.6$~keV in the SIS.
The open stars show the corresponding data/model ratio 
(again rebinned for clarity)
when this model is {\it directly compared} to the 
{\it ROSAT} PSPC data.}
\end{figure}
\clearpage

\begin{figure}
\plotfiddle{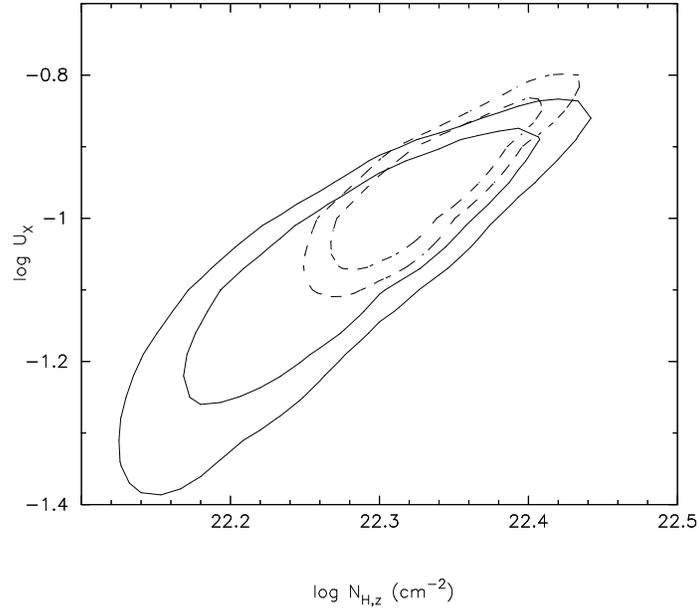}{8cm}{0}{50}{50}{-140}{0}
\caption{
$\chi^2$ contours corresponding to the 68 and 95\% confidence regions 
(for 3 interesting parameters: $\Gamma$, $U_X$ \& $N_{H,z}$) 
from the {\it ASCA} analysis alone (solid curves)
and the analysis of the joint {\it ASCA}/{\it ROSAT} analysis
(dashed curves). 
Clearly both analyses are consistent with an 
X-ray ionization parameter $U_X \simeq 0.1$ and 
column density $N_{H,z} \simeq 2\times10^{22}\ {\rm cm^{-2}}$.}
\end{figure}
\clearpage

\begin{figure}
\plottwo{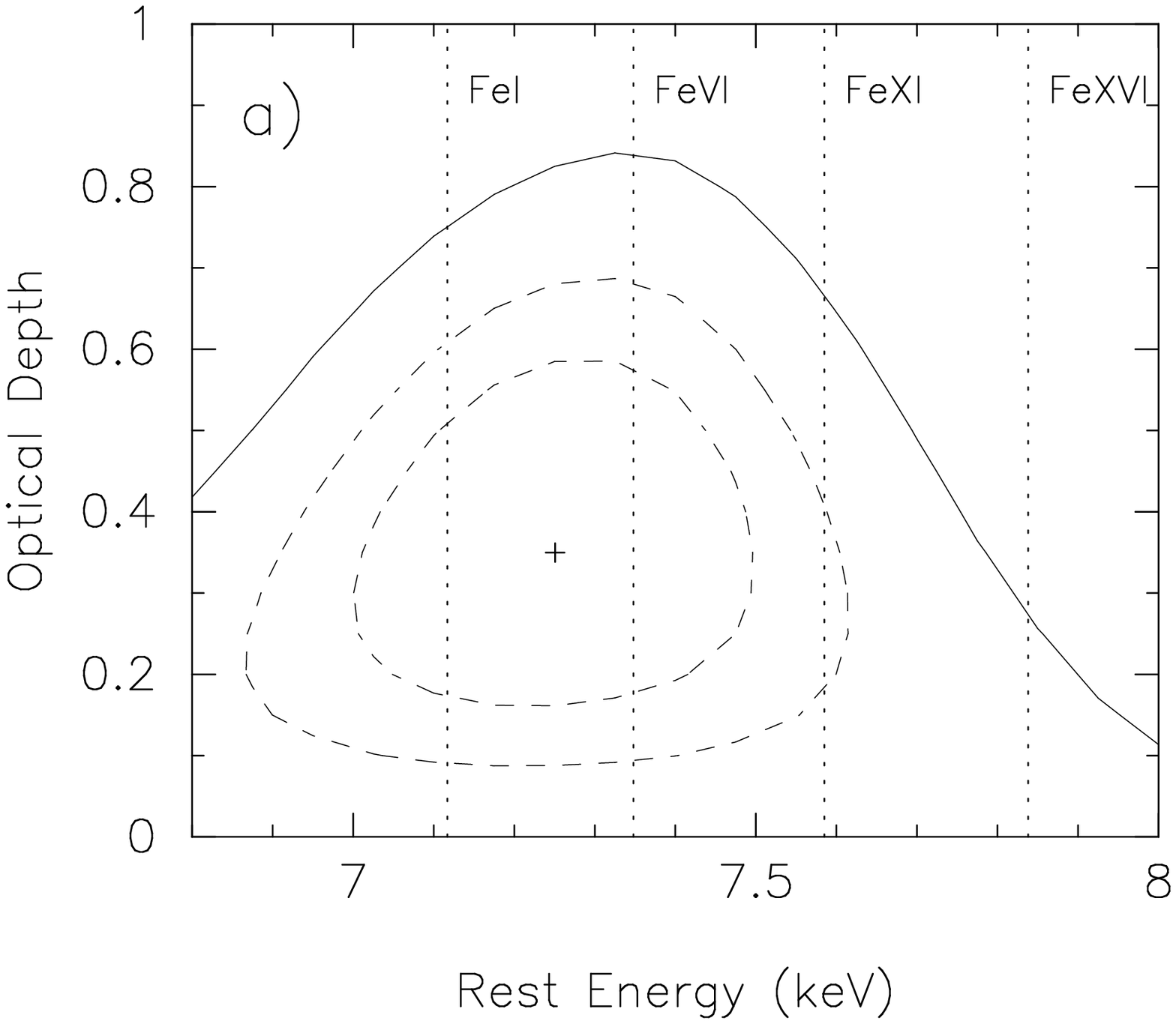}{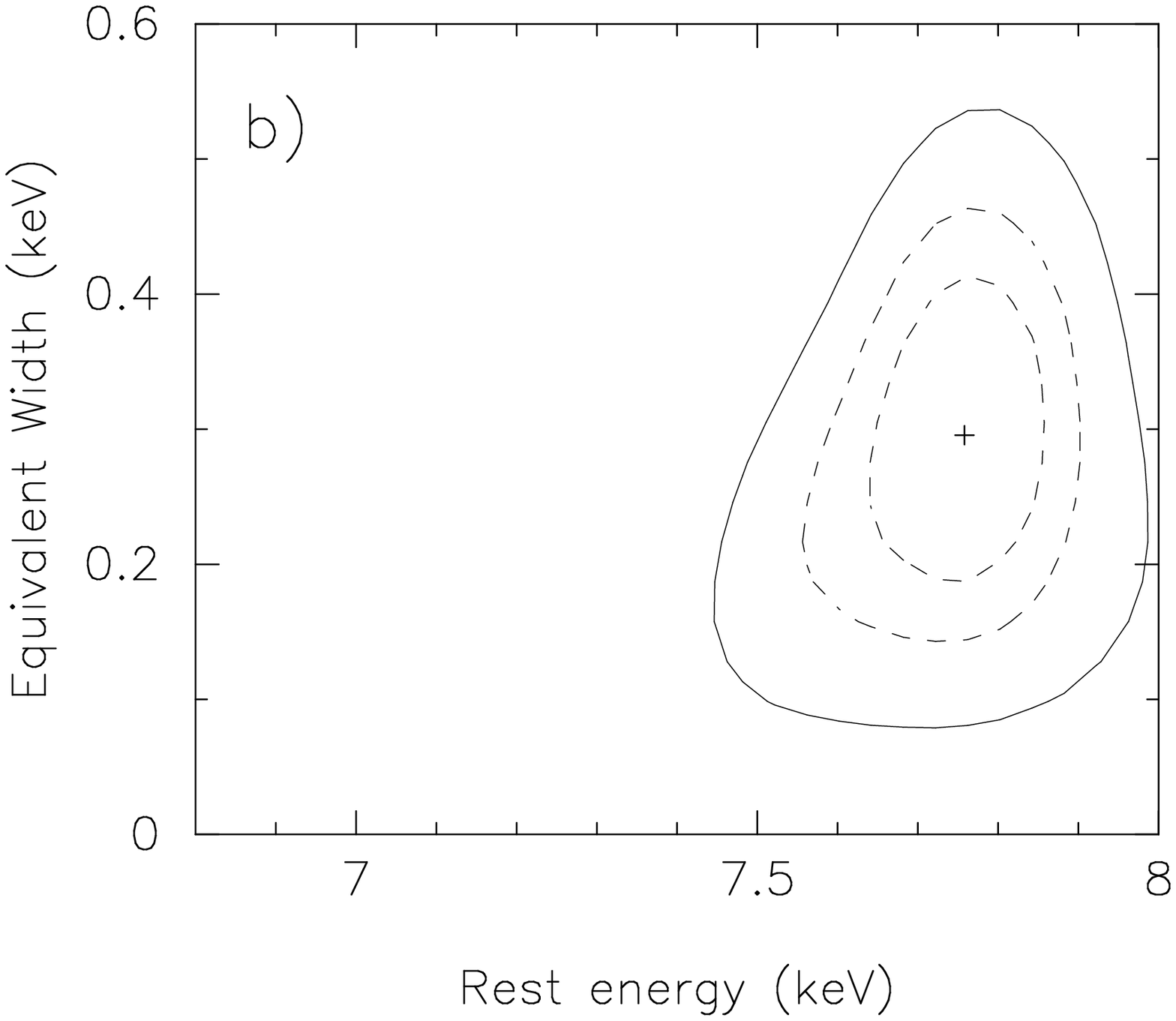}
\caption{
a) The dashed curves show the $\chi^2$ contours corresponding to the
68 and 90\% confidence regions 
for 2 interesting parameters ($E_{z}^{abs}$ \& $\tau$), 
when an additional absorption 
edge is added to ionized-absorber model described in \S\ref{Sec:wabs}.
The solid curve shows the contour corresponding to
90\% confidence for
5 interesting parameters
($\Gamma$, $U_X$, $N_{H,z}$, $E_{z}^{abs}$ \& $\tau$). 
As can be seen, such a feature is consistent with a variety of 
ionization states of iron up to $\sim$Fe{\sc xix}.
b) As above, but for when the 'notch'-shaped absorption feature 
is added to ionized-absorber model. 
Such a feature is consistent with 
resonance scattering by Ni $K\alpha$ ($<$Ni{\sc xxviii})
and Fe $K\beta$ ($>$Fe{\sc xx}) 
in the rest--frame of the quasar
or 
Fe $K\alpha$ in out-flowing material (see \S\ref{Sec:fe-band}).}
\clearpage
\end{figure}
\clearpage

\end{document}